\documentclass[twocolumn,showpacs,preprintnumbers,nofootinbib,prl,superscriptaddress,groupedaddress,10pt]{revtex4-1}

\pdfoutput=1

\usepackage{graphicx,amssymb,amsmath,amsthm,amsfonts,epsfig,epsf}
\usepackage[linktocpage]{hyperref}
\usepackage[usenames]{color}
\usepackage{epstopdf}
\usepackage{appendix}
\usepackage{aas_macros}
\usepackage{bm}
\usepackage{dcolumn}
\usepackage[latin1]{inputenc}
\usepackage{latexsym}
\usepackage{rotating}
\usepackage{hyperref}
\usepackage{color}
\usepackage{longtable}
\usepackage{enumerate}
\usepackage{tensor}
\usepackage{mathtools}
\usepackage{url}
\renewcommand{\vec}[1]{\boldsymbol{#1}}

\newcommand{\ben}{\begin{enumerate}}
\newcommand{\een}{\end{enumerate}}

\def\be{\begin{equation}}
\def\ee{\end{equation}}
\newcommand{\beq}{\begin{eqnarray}}
\newcommand{\eeq}{\end{eqnarray}} 
\newcommand{\ba}{\begin{align}}
\newcommand{\ea}{\end{align}}
\newcommand{\bsub}{\begin{subequations} \begin{eqnarray}}
\newcommand{\esub}{\end{eqnarray} \end{subequations}}

\newcommand{\p}{\partial}

\newcommand{\om}{\omega}
\newcommand{\Om}{\Omega}
\newcommand{\tom}{\widetilde \om}
\newcommand{\sig}{\sigma}

\begin{document}

\title{Detecting Rotational Superradiance in Fluid Laboratories}

\author{
Vitor Cardoso$^{1,2,3}$ 
,
Antonin Coutant$^{4}$
,
Mauricio Richartz$^{5}$
,
Silke Weinfurtner$^{4}$
}
\affiliation{${^1}$ CENTRA, Departamento de F\'{\i}sica, Instituto Superior T\'ecnico -- IST, Universidade de Lisboa -- UL,
Avenida Rovisco Pais 1, 1049 Lisboa, Portugal}
\affiliation{${^2}$ Perimeter Institute for Theoretical Physics Waterloo, Ontario N2J 2W9, Canada}
\affiliation{${^3}$ Theoretical Physics Department, CERN, CH-1211 Gen\`eve 23, Switzerland}
\affiliation{${^4}$ School of Mathematical Sciences, University of Nottingham, University Park, Nottingham, NG7 2RD, UK}
\affiliation{${^5}$ Centro de Matem\'atica, Computa\c c\~ao e Cogni\c c\~ao,
Universidade Federal do ABC (UFABC), 09210-170 Santo Andr\'e, SP, Brazil.
}


\begin{abstract}
Rotational superradiance was predicted theoretically decades ago, and is chiefly responsible for a number of important effects and phenomenology in black hole physics.
However, rotational superradiance has never been observed experimentally. Here, with the aim of probing superradiance in the lab, we
investigate the behavior of sound and surface waves in fluids resting in a circular basin at the center of which a rotating cylinder is placed. 
We show that with a suitable choice for the material of the cylinder, surface and sound waves are amplified. Two types of instabilities are studied: one sets in whenever superradiant modes are confined near the rotating cylinder; the other, which does not rely on confinement, corresponds to a local excitation of the cylinder. Our findings are experimentally testable in existing fluid laboratories and hence offer experimental exploration and comparison of dynamical instabilities arising from rapidly rotating boundary layers in astrophysical as well as in fluid dynamical systems. 
\end{abstract}

\maketitle


\emph{\textbf{Introduction.---}} Superradiant amplification of waves by rotating systems has been predicted to arise in a variety of setups~\cite{Brito:2015oca}. 
Zel'dovich pioneered the studies of superradiance by rotating objects, showing that low-frequency electromagnetic waves scattered by a rotating conducting cylinder are amplified~\cite{zeldovich1,zeldovich2,zeldovich3}. Following the original proposal, Misner~\cite{misner} suggested that rotating black holes would also amplify low-frequency waves, which was confirmed analytically and numerically
by several authors~\cite{staro1,staro2,Teukolsky:1974yv,Brito:2015oca}. More generally, in curved spacetimes, superradiance leads to a wealth of interesting phenomenology, including 
floating orbits~\cite{Cardoso:2011xi} and superradiant instabilities which lead to interesting constraints on ultralight fields~\cite{Arvanitaki:2010sy,Pani:2012vp,Brito:2013wya} and even new hairy black hole configurations~\cite{Herdeiro:2014goa}. Upon quantization, superradiance yields spontaneous emission of radiation, which was historically a precursor to black hole evaporation.  

Despite the wealth of theoretical and astrophysical implications, rotational superradiance has never been observed.
In this letter, we discuss rotational superradiance in fluids, with the aim of implementing such setup in the laboratory. Our analysis builds on previous work, which showed the existence of superradiant scattering of both sound and surface waves by a rotating analogue black hole~\cite{basak, basak2, Schutzhold:2002rf,Berti:2004ju,Richartz:2014lda} (the main difference in comparison to our work is the energy extraction mechanism, which relies on the trapping of modes inside the analogue event horizon). In fluid dynamics, this amplification effect is called overreflection~\cite{booker, mckenzie, acheson, kelley} and has been studied mostly for vortex sheets in supercritical flows (flow velocities larger than the wave velocity).
Our main objective is to investigate the possibility of superradiant amplification and superradiant instabilities for waves propagating on a static fluid flow around a central rotating cylinder. 
The cylinder is characterized by an impedance, which is a well known concept in acoustics, describing the interaction between the wave and the scatterer. Several factors like thickness, porosity and fiber size influence its value~\cite{seddeq2009}. The possibility of manipulating the impedance of a given object~\cite{quan2014,Yang} motivates and strengthens our analysis. For this reason, we will provide a precise account of how the value of the impedance influences superradiance.

\emph{\textbf{The setup.---}} Consider a (initially static) fluid of density $\rho_0$ constrained between two concentric cylinders of \textit{radii} $R_0$ and $R_1>R_0$. The inner cylinder rotates with constant angular velocity $\Omega$, whereas the outer cylindrical wall is at rest. After one sets the cylinder in motion, the initially still fluid will be dragged by the cylinder's motion until it reaches an equilibrium state. This is the circular Couette flow. We assume that $\Omega$ is sufficiently small to avoid the formation of Taylor vortices~\cite{taylorcouette1,taylorcouette2}. As discussed in the Supplemental Material, all viscous effects can be neglected~\cite{misc}. \nocite{Landau:book_fluids,Johnson} In particular, the time scales associated with superradiant effects are much lower than the time scale of the diffusion of angular momentum from the surface of the cylinder to the flow. For all practical purposes the fluid can be considered to be at rest.

\emph{\textbf{Scattering off a rotating cylinder.---}} The propagation of both sound and surface waves on a static and inviscid flow is described by a scalar field $\psi_1(t,r,\phi) = (\varphi(r)/\sqrt{r})e^{im \phi - i \omega t}$ which obeys the wave equation
\be 
\p_r^2 \varphi(r) + \left[\frac{\omega^2}{c^2} - \frac{1}{r^2}\left(m^2 - \frac{1}{4}\right) \right] \varphi(r) = 0.  \label{radialsound} 
\ee 
Here $\omega$ is the frequency of the wave and $m$ is the azimuthal wave number. 
 
We assume that the outer cylinder is sufficiently far away from the inner one that its presence can be ignored in a scattering experiment. Far away from the inner cylinder ($r \gg R_0 $), the wave is composed by an incident part with amplitude $|\mathcal{A}^{\scriptstyle -}|$ and a reflected part with amplitude $|\mathcal{A}^{\scriptstyle +}|$, that is \be \label{radial_as}
\left. \varphi(r) \right\vert_{r\rightarrow \infty}= \mathcal{A}^{\scriptstyle -} e^{-i \om r/c} + \mathcal{A}^{\scriptstyle +} e^{i \om r/c}. 
\ee

At the surface of the cylinder, waves can exchange energy and momentum with the cylinder. This process is encoded in a boundary condition involving the impedance $Z_\om$ of the cylinder, which is a complex number relating the pressure change to the radial velocity perturbation on the cylinder~\cite{Rienstra}. In general, it depends on the frequency of the perturbation~\cite{delany}. 

The real part of the impedance is called resistance and relates to the energy flow. We assume $\mathrm{Re}(Z_\om) > 0$, corresponding to a surface which absorbs energy.
The imaginary part, on the other hand, is called reactance and relates to the natural oscillation frequency of the surface. If the frequency of the incident wave resonates with the surface, one has $\mathrm{Im}(Z_\om) \sim 0$, as in~\cite{aso}. On the other hand, far from resonance, the surface of the cylinder barely moves under the action of the wave and behaves like a hard wall. 

In terms of impedance, the boundary condition for both sound and surface waves in the rest frame of the cylinder is   
\be \label{BCcyl0}
\left. \left( \frac{\p_r \psi_1}{\psi_1} \right) \right|_{r=R_0} =  - \frac{i \rho_0 \om}{Z_{\om}}.
\ee
When the cylinder rotates uniformly with angular velocity $\Omega$, it is sufficient to transform to a new angular coordinate $\widetilde \phi=\phi+\Omega t$, which effectively amounts to the replacement of $\omega$ with $\tom = \omega-m\Omega$ in \eqref{BCcyl0}. In terms of the radial field $\varphi$, we have
 \be \label{BCcyl}
\left. \p_r \varphi \right|_{r=R_0} = \left(\frac{1}{2R_0} - i \rho_0 \frac{\tom}{Z_{\tom}} \right) \left. \varphi \right|_{r=
R_0}.
\ee

\emph{Superradiance.---} Eq.~\eqref{radialsound} admits the general solution 
\begin{equation} \label{Eq:GenSolution}
\varphi(r)=C_1\sqrt{r}\,J_m(\omega r/c)+C_2\sqrt{r}\,Y_m(\omega r/c) \,, 
\end{equation}
where $J_m$ and $Y_m$ are the Bessel functions of first and second kinds. Because of the boundary condition \eqref{BCcyl}, the constants $C_1$ and $C_2$ are related to the scattering coefficients $\mathcal{A}^{\scriptstyle \pm}$ by 
\be
\mathcal{A}^{\scriptstyle \pm} = \sqrt{\frac{c}{2\pi\omega}}(C_1 \mp iC_2)e^{\mp i(m\pi/2+\pi/4)}.
\ee

To characterize superradiance, it is common practice to exploit the Wronskian $W = \mathrm{Im}(\varphi^* \p_r \varphi)$, 
which
is independent of $r$~\cite{Richartz:2009mi,Brito:2015oca}. At large $r$, it is 
\be
W = \frac{\om}{c} (|\mathcal{A}^{\scriptstyle +}|^2 - |\mathcal{A}^{\scriptstyle -}|^2) = \frac{\om}{c} A_{\omega m} |\mathcal{A}^{\scriptstyle -}|^2.
\ee
The number $A_{\omega m}=|\mathcal{A}^{\scriptstyle +}|^2/|\mathcal{A}^{\scriptstyle -}|^2-1$ is defined as the amplification factor. When $A_{\omega m} > 0$, the scattering process is referred to as superradiance since the amplitude of the reflected wave is larger than the amplitude of the incident one. 
By equating the Wronskian above with the Wronskian at the surface of the cylinder, one can relate the amplification factor with the impedance: 
\be \label{wronskian}
A_{\omega m} |\mathcal{A}^{-}|^2 = - \frac{\rho_0 c \, \tom \mathrm{Re}(Z_{\tom})}{\om |Z_{\tom}|^2} |\varphi(R_0)|^2. 
\ee
As we see, the occurrence of superradiance depends solely on the signs of $\tom$ and $\mathrm{Re}(Z_{\tom})$. As explained before, the real part of the impedance is positive and, therefore, superradiance will occur as long as $0 < \om < m\Om$, which is the usual superradiant condition for waves scattering off a rotating object~\cite{Brito:2015oca}.\\

\emph{Amplification coefficients.---}
 Unfortunately, Eq.~\eqref{wronskian} is not useful for determining the numerical value of the amplification factor; the full solution of the wave equation is needed. Combining Eqs.~\eqref{BCcyl} and \eqref{Eq:GenSolution}, one finds
\be \label{sigmasound}
A_{\omega m} = \left|\frac{\frac{\sig-1}{\sig} Y_m - i  Z Y_m'+ i\frac{\sig - 1}{\sig} J_m + Z J_m'}{\frac{\sig-1}{\sig} Y_m - i Z  Y_m'- i \frac{\sig-1}{\sig} J_m -  Z J_m'}\right|^2 -1,
\ee
where the dimensionless parameters $\sigma = \omega / (m\Om)$,
$Z=Z_{\tom}/(\rho_0 c)$, and $\alpha= \Omega R_0/c$ have been defined.
For short, we write $J_i=J_i(m \alpha \sig)$ and $Y_i=Y_i(m \alpha \sigma)$. Note that $\alpha$ measures how fast the cylinder is rotating: $\alpha < 1$ ($\alpha>1$) represents a subcritical (supercritical) cylinder, which rotates slower (faster) than the wave speed.  

\begin{figure}[t]
\begin{center}
\begin{tabular}{c}
\includegraphics[width=7.0cm]{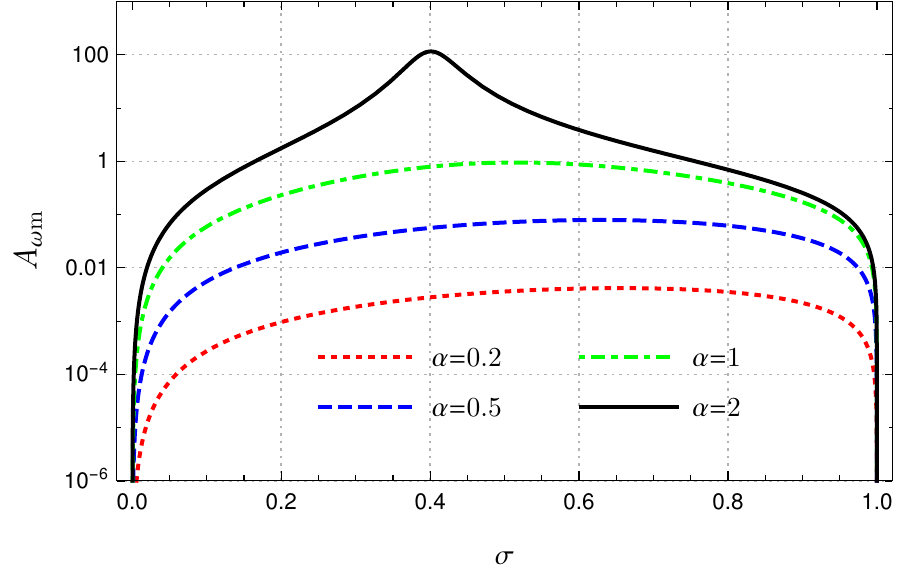}
\end{tabular}
\caption{Amplification factors $A_{\omega m}$ for $Z= 1-i$ and different rotation speeds.
\label{fig2}
}
\end{center}
\end{figure}

In Fig.~\ref{fig2} we plot the amplification factor as a function of $\sig$ for different rotation speeds and a fixed impedance $Z= 1 - i$, which is a typical value for the impedance of a fibrous material~\cite{delany}. Here and throughout, we assume $m=1$ since it maximizes the amplification for most values of $Z$.
For fixed $\alpha$, $A_{\om m}$ attains its maximum $A_{ m}^{\mathrm{max}}$ at the superradiant frequency $\sig _{\mathrm{max}}$.

\begin{figure}[t]
\begin{center}
\begin{tabular}{c}
\includegraphics[width=7.0cm]{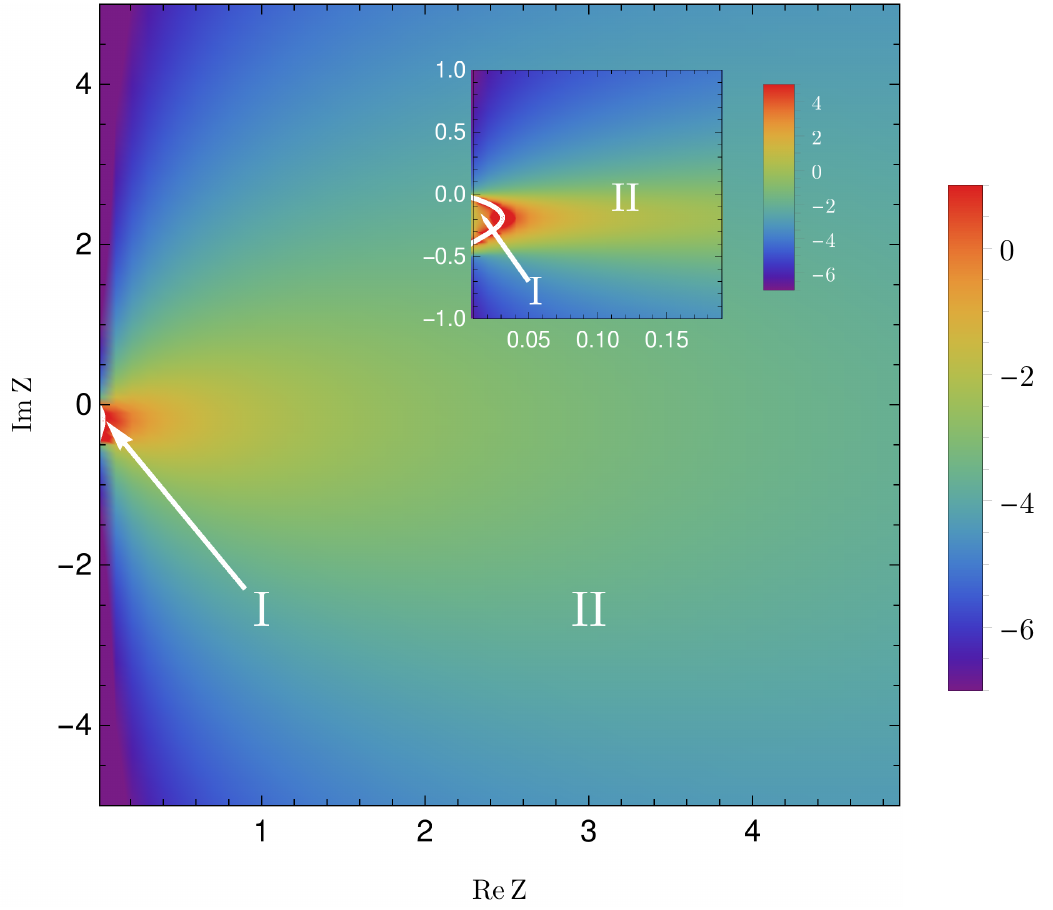} \\
\includegraphics[width=7.0cm]{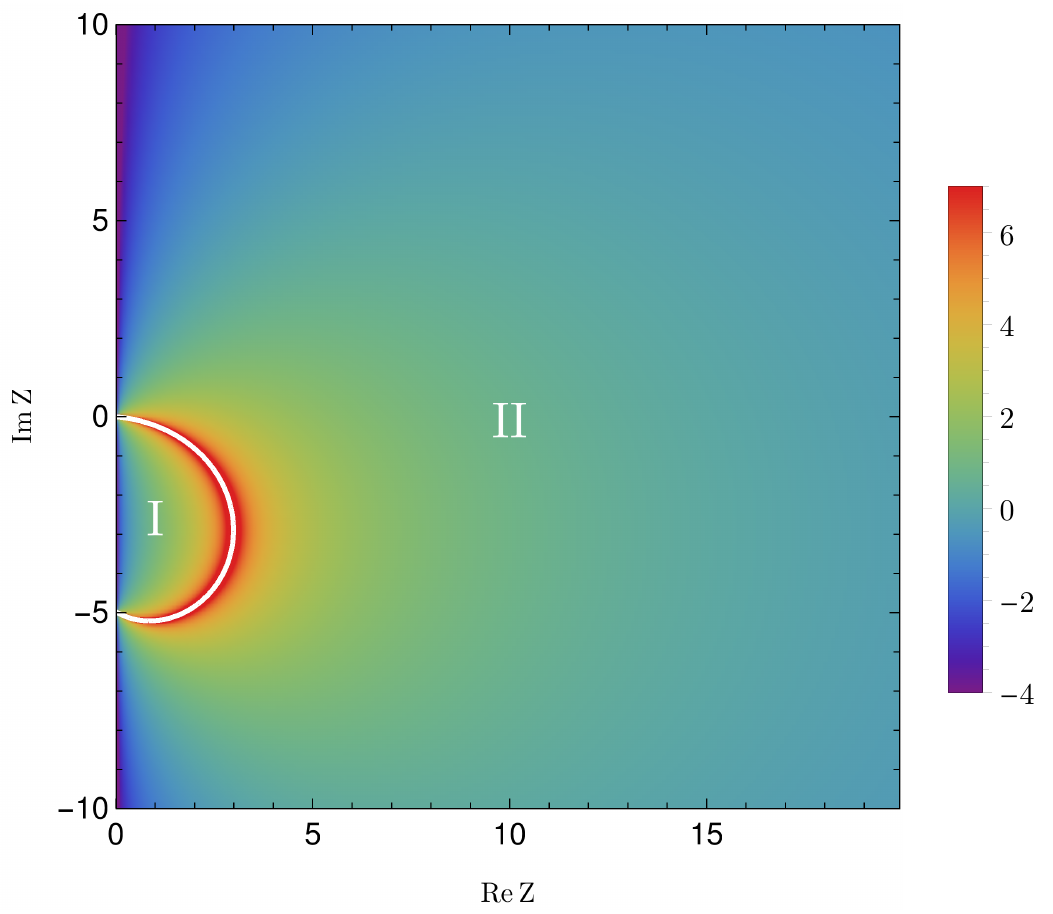} 
\end{tabular}
\caption{Logarithm of the maximum amplification $\log(A_{m} ^{\mathrm{\max}})$ as a function of the real and imaginary parts of the impedance for $\alpha=0.5$ (top panel) and $\alpha=5$ (bottom panel). The white curve separates the complex plane according to the stability of the system: region I (inside) is unstable while region II (outside) is stable.
\label{fig34}
}
\end{center}
\end{figure}

To have a better understanding on how large $A_{ m}^{\mathrm{max}}$ can be, we analyze the limits of small and large wavelengths 
$\lambda$ using the relation $\lambda/R_0 = 2 \pi / (m \alpha \sig)$. In fact, if the radius of the inner cylinder is much smaller than the wavelength $\lambda$ of the incident waves, i.e.~$ m \alpha\sig \ll 1$, the amplification factor (for $m>0$) reduces to  
\be \label{eqamp1}
A_{\omega m} = 
 -\frac{8 \pi  \mathrm{Re}(Z) (\sig - 1)}{2^{2m} (m!)^2 |Z|^2 \sig } \left(m \alpha \sig \right)^{2 m+1}.
\ee
This formula holds both in the sub- and supercritical regimes. Because of the power law decay, the resulting amplification factor is typically very small ($A_{\om m} \ll 1$) in this limit, unless the impedance is small $|Z| \sim 0$ and/or the cylinder is fast ($\alpha \gg 1$).

On the other hand, if $ m \alpha \sig \gg 1$, corresponding to $\lambda / R_0 \ll 1$, the amplification factor (for $m\ge 0$) can be recast as
\be \label{eqamp2}
A_{\omega m} = 
 -\frac{4 \mathrm{Re}(Z) \sig (\sig - 1)}{ [(1+\mathrm{Re}(Z))\sig - 1]^2 + \mathrm{Im}(Z)^2\sig ^2},
\ee
which does not depend on $\alpha$.
To obtain a significant amplification in this situation, one needs a material with $\mathrm{Im}(Z)\approx 0$. Indeed, if $\kappa = \mathrm{Re}(Z)/ \mathrm{Im}(Z)$, the maximum amplification is 
$A_{m} ^{\mathrm{\max}} = 2 \kappa ( \kappa + \sqrt{\kappa ^2 + 1})$ 
and occurs at $\sig_{\max} =  (1+|Z|)^{-1}$.
In this case the small wavelength assumption is equivalent to $\Omega R_0 \gg c(1+|Z|)$ and, therefore, the cylinder must be supercritical.

It is natural for the amplification factor to depend on the impedance and on the velocity of the cylinder. To illustrate, we plot in Fig.~\ref{fig34} $\log(A_{m} ^{\mathrm{\max}})$ 
as a function of $\mathrm{Re}(Z)$ and $\mathrm{Im}(Z)$ for $\alpha=0.5$ and $\alpha=5$. We observe that there is no significant difference in the qualitative behavior of $A_{m} ^{\mathrm{\max}}$ when varying the rotation speed (i.e., $\alpha$), except for a change in the scale of $Z$. In particular, superradiance occurs for both sub- and supercritical cylinders, as one can compensate for a lower angular velocity with a lower impedance. 
More generally, any value of the amplification can be obtained at fixed $\alpha$, by properly choosing the impedance. 
Arbitrarily small values of $A_{m} ^{\mathrm{\max}}$  
are obtained as $|Z|$ is increased. Arbitrarily large values of $A_{m} ^{\mathrm{\max}}$, on the other hand, are obtained for impedances which lie close to the white curves in Fig.~\ref{fig34}.

These white curves represent points where $A_{\om m}$ diverges. They correspond to solutions of the wave equation which are purely outgoing far away from the cylinder, i.e.~for which $\mathcal{A}^{-}$ in Eq.~\eqref{radial_as} vanishes. This extra boundary condition turns the scattering problem into an eigenvalue problem for the frequency, and is equivalent to finding the poles $\sig = \sig_R + i \sig_I$ of $A_{\om m}$ in the complex plane. 
By looking at the corresponding equation, we can show that there is a unique solution for each $m$.  
If $\sig_I < 0$, the corresponding mode decays in time (analogously to the ringdown of a black hole~\cite{Berti:2009kk}). On the other hand, if $\sig_I > 0$, the mode is unstable, meaning that it will grow in time until the linear approximation breaks down. The white lines correspond to an eigenmode whose imaginary part is exactly zero. They divide the $Z$-complex right halfplane into two regions (see Fig.~\ref{fig34}): inside (region I) an unstable mode exists; outside (region II) the mode is stable.

This unstable mode corresponds to a local excitation of the surface of the cylinder. Indeed, for a negative reactance $\mathrm{Im}(Z) < 0$, there exist surface waves that are evanescent in the radial direction~\cite{Rienstra}. When including rotation, such a mode possesses a negative energy when it lies inside the superradiant regime $0 < \om < m \Om$. Since it is evanescent in the radial direction, it couples with the continuum of modes in the fluid through a mechanism analogous to tunneling. This leads to the observed dynamical instability: the negative energy modes are amplified by emitting positive energy waves in the fluid~\cite{Coutant:2016bgk}.In Fig.~\ref{fig34}, the stable and unstable regions are plotted for $m=1$. When considering higher values of m, these regions tend to increase or decrease, depending on $\alpha$ (but the corresponding instability time scale does not change much).

\begin{figure}[t]
\begin{center}
\begin{tabular}{c}
\includegraphics[width=7cm]{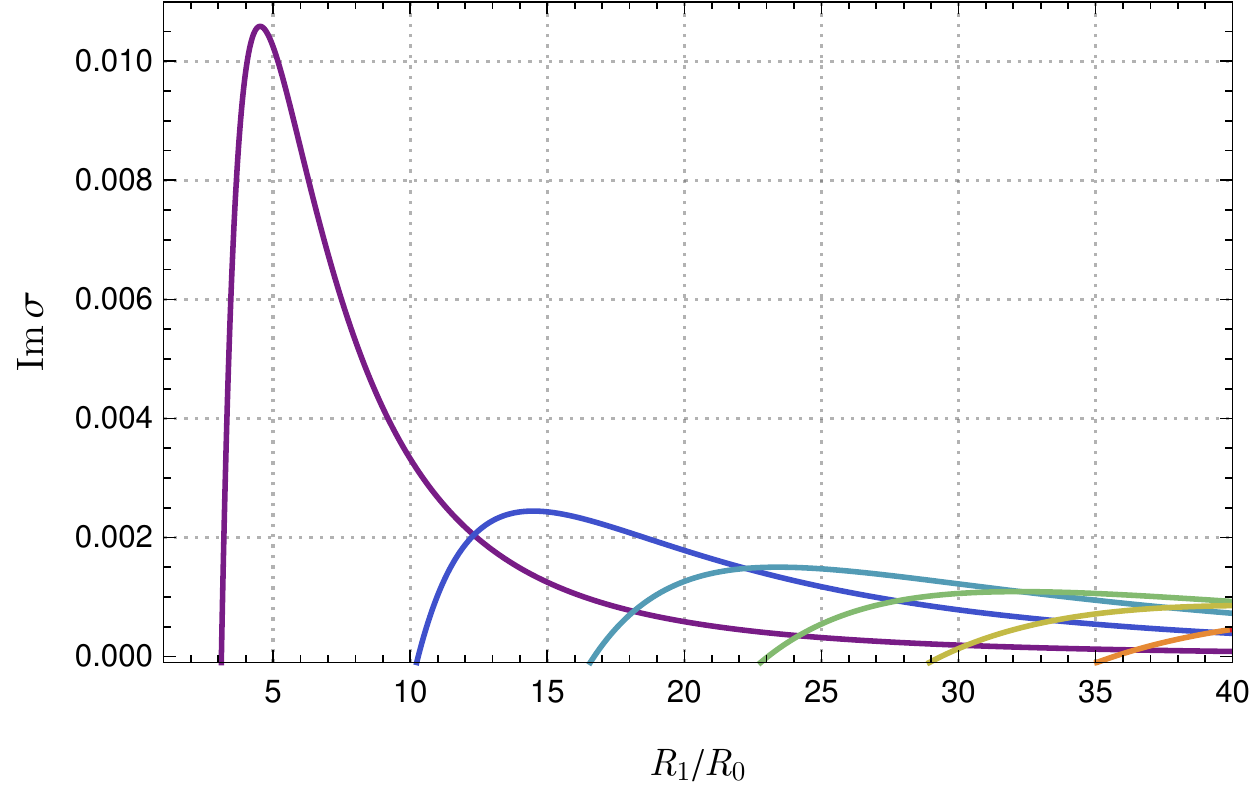} \\
\includegraphics[width=7cm]{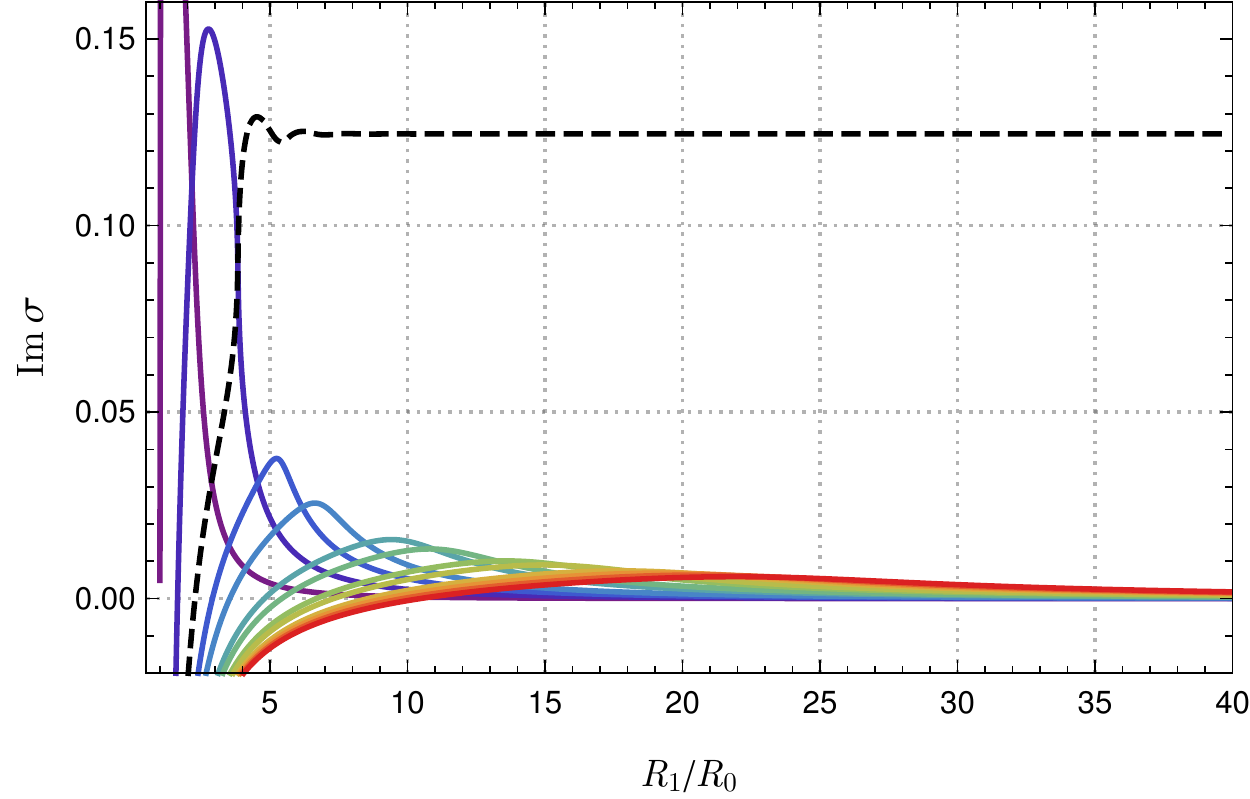}
\end{tabular}
\caption{Instability rate for the ``acoustic bomb" modes as a function of the ratio $R_1/R_0$ for $\alpha = 0.5$ (top panel) and $\alpha = 5$ (bottom panel). The impedance considered is $Z=1- 1 \, i$. 
\label{fig56}
}
\end{center}
\end{figure}
\begin{figure}[t]
\begin{center}
\begin{tabular}{c}
\includegraphics[width=7cm]{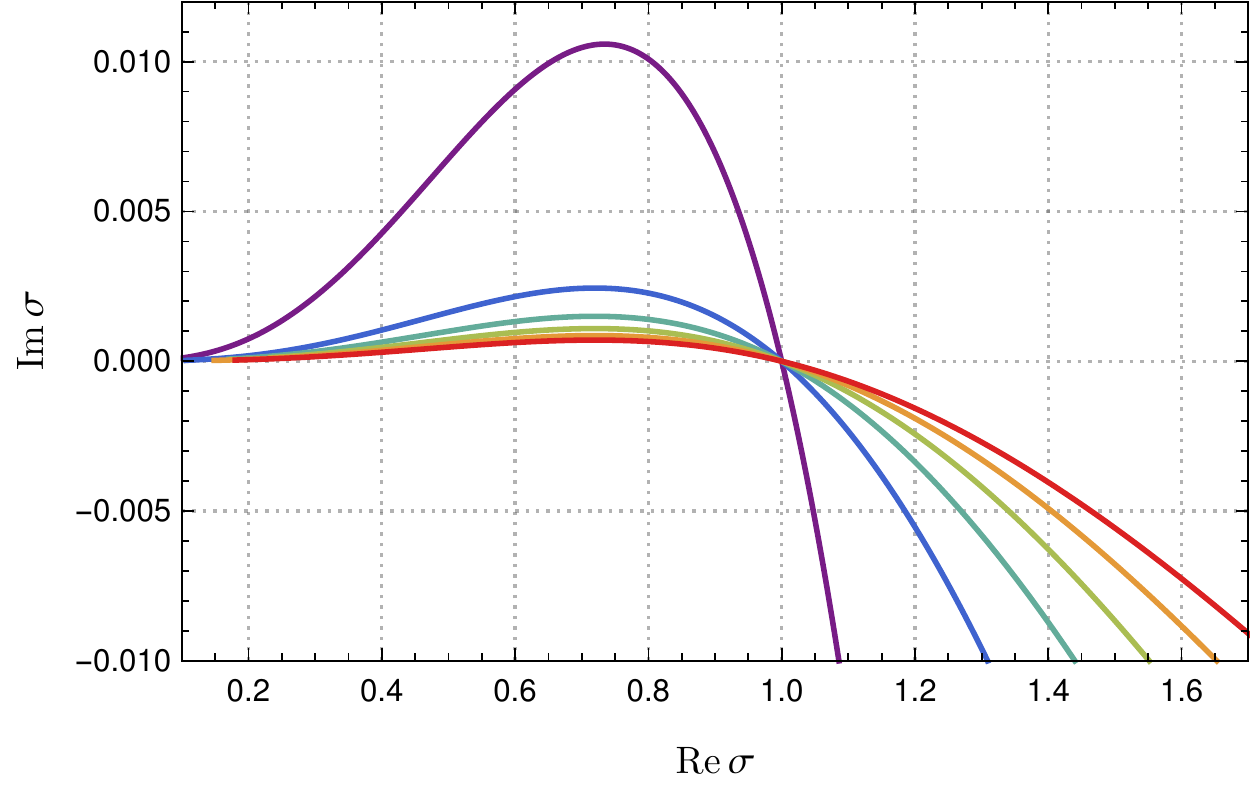} \\
\includegraphics[width=7cm]{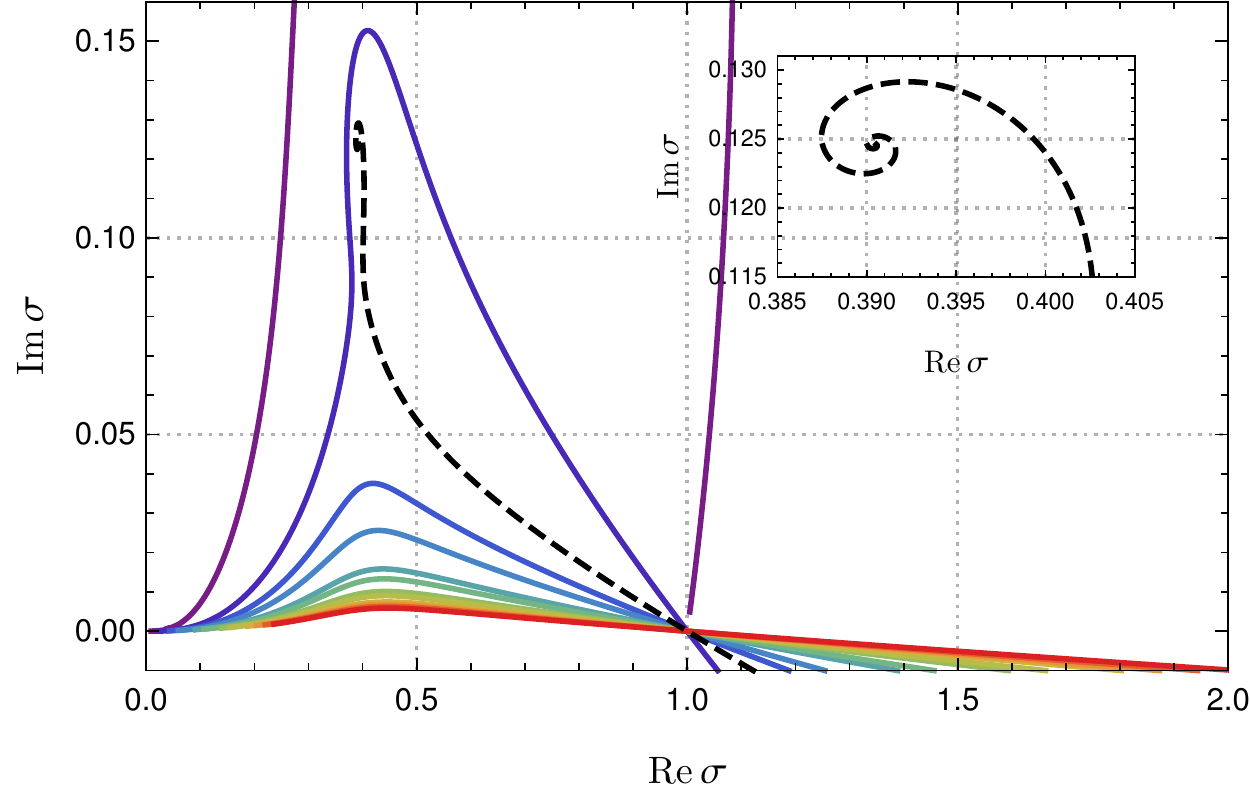}
\end{tabular}
\caption{Parametric plot of the instability rate and the oscillation frequency of the modes confined between the cylinders. The parameter used is the ratio $R_1/R_0$. (Inset) The peculiar mode which spirals towards an unstable eigenvalue as $R_1/R_0$ increases is shown in detail.
\label{fig78}
}
\end{center}
\end{figure}

\emph{\textbf{Superradiant instabilities: ``Acoustic bomb".---}}  Another interesting application of superradiance is the possibility to build an ``acoustic bomb''. By confining the superradiant modes near the rotating cylinder we can amplify the superradiant extraction of energy and trigger another kind of instability. In this simple setup, confinement is achieved by the presence of the outer cylindrical surface (which is characterized by an impedance $Z^{\mathrm{ext}}$). Unlike the instability discussed in the previous section, this one occurs for any impedance. 
After imposing the appropriate boundary conditions, we obtain from \eqref{Eq:GenSolution} the equation for the associated eigenfrequencies,
\be \label{Eq:bhbomb}
\frac{i (\sig - 1) J_m + \sig Z J_m'}{i (\sig - 1) Y_m + \sig Z Y_m'} = \frac{i\widehat{J}_m -  Z ^{\mathrm{ext}} \widehat{J}_m'}{i\widehat{Y}_m -  Z ^{\mathrm{ext}} \widehat{Y}_m'},
\ee
where we have further defined $\widehat{J}_m=J_m(m\alpha \sigma R_{1}/R_0)$ and $\widehat{Y}_m=Y_m(m\alpha \sigma R_{1}/R_0)$ for short.   

As a point of principle, we solve Eq.~\eqref{Eq:bhbomb} by taking $Z\sim 1-i$ and assuming a ``perfect mirror'' ($|Z^{\mathrm{ext}}|\rightarrow\infty$) configuration, which is akin to the model proposed by Press and Teukolsky for the ``black hole bomb"~\cite{Press:1972zz,Cardoso:2004nk,Cardoso:2004hs}. 
As the distance between the inner and outer cylinders increase, we show in Fig.~\ref{fig56} that the ``acoustic bomb" instability rates typically decrease (note that for a given ratio $R_1/R_0$, several unstable modes coexist). 

In Fig.~\ref{fig78} the real and imaginary parts of the eigenfrequencies are plotted using the ratio $R_{1}/R_0$ as a parameter. Note that the unstable modes usually satisfy the superradiance condition $0 <\sig_R< 1$. A remarkable fact regarding Fig.~\ref{fig56} is the existence of a peculiar mode (for $\alpha=5$) whose instability rate tends to a non-zero value as the ratio $R_1/R_0$ increases. This unstable mode is 
simply the unstable mode described in the preceding section, only slightly perturbed by the outer boundary (when $R_1/R_0 \rightarrow \infty$, it becomes a pole of $A_{\om m}$). The corresponding spiraling behavior of the frequency in the complex plane resembles the behavior of some quasinormal modes of near extremal rotating black holes~\cite{onozawa2,branching2,richartzqne}.

  For the subcritical cylinder ($\alpha = 0.5$), the fastest growing mode is $\sig \sim 0.75 + 0.01 \,i $ and occurs when $R_{1}/R_0 \sim  4.5$. The associated instability time scale is $\tau_i = 100/ \Omega$. For the supercritical cylinder ($\alpha = 5$), on the other hand, if $R_1/R_0 \gtrsim 5$, the characteristic time of the instability is $\tau_i = 8/ \Omega$, which is even shorter than the previous result. 

\emph{\textbf{Final Remarks.---}} Although our model is extremely simple, the results we have obtained suggest the interesting prospect of detecting superradiant amplification and ``acoustic bomb'' instabilities in a fluid laboratory. We offer an alternative to existing proposals involving analogue black holes (see  \cite{Faccio} for another recent proposal, involving nonlinear optics).
The main message of this letter is that, by carefully selecting the impedance of a rotating cylinder, one can observe superradiant effects. In particular, although amplification increases with angular velocity, it is not necessary for the cylinder to rotate at a speed that surpasses the wave speed. At low rotation speeds, high amplifications can be obtained by using a cylinder with a small impedance. Because the impedance is a function of the frequency, in order to observe the effect one has to match the required impedance for large superradiance with the experimental frequency. 

Concerning the experimental observation of superradiance for surface waves in today's available fluid laboratories, a reasonable estimate for the parameters of the setup is: $R_0 \sim 0 - 0.2 \, {\rm m}$, $R_1 \sim 0.5 - 2 \, {\rm m}$, $h_0 \sim 0 - 0.5 \, {\rm m}$, $\om \sim 0 - 10 \, {\rm s}^{-1}$, and $\Om \sim 0 - 10 \, {\rm s}^{-1}$. The corresponding velocity of the inner cylinder is $\Om R_0 \sim  0 - 2 \, {\rm m }\, {\rm s}^{-1}$. By adjusting the fluid depth, one can set the wave speed to be as large as $c \sim 2 \, {\rm m} \, {\rm s}^{-1}$. Hence, both sub- and supercritical velocities are possible in a realistic experiment. In particular, both cases discussed in this letter ($\alpha=0.5$ and $\alpha=5$) are reproducible for surface waves. If an inner cylinder with the appropriate material (impedance) can be designed, superradiant effects should be detectable in the laboratory.  

For sound waves, on the other hand, the velocities are typically much larger ($c \sim 1500 \, {\rm m} \, {\rm s}^{-1}$ in water and $c \sim 300 \, {\rm m} \, {\rm s}^{-1}$ in air). Hence much larger rotation speeds, wave frequencies and cylinder radii are required for experimental implementation. For a given rotation speed $\Om$, the parameter $\alpha$ for surface waves is typically two or three orders of magnitude larger than the $\alpha$ parameters for sound waves. Conversely, for a given cylinder with impedance $Z_{\tom}$, the dimensionless parameter $Z$ is two or three orders of magnitude larger for surface waves than for sound waves. Taking all the details into account, we believe surface waves provide the easiest setup for experimental verification of superradiance. Nonetheless, the techniques recently developed in the field of acoustic metamaterials~\cite{quan2014, Yang} offer very promising directions for sound waves as well.

In summary, although superradiant scattering has been know for over 40 years, it has never been observed. We propose here a feasible experimental setup inspired by Zel'dovich's cylinder and by rotating black holes. We provide a detailed analysis of the influence of the impedance on the scattering amplitudes and show that, by carefully choosing it, arbitrarily large amplification coefficients can be obtained. On the contrary, the gain coefficient in the original superradiance proposal by Zel'dovich is extremely small unless the cylinder's velocity is comparable to the speed of light~\cite{bek_1}. Unlike the case for analogue rotating black holes, where one has very little control over the background flow, in our case the flow is much simpler (static) and everything is encoded in the impedance. The possibility to better control the amplification through the impedance offers an excellent opportunity to observe superradiance, in either sound or surface waves.

\emph{\textbf{Acknowledgements.---}} V.C.~acknowledges financial support provided under the European
Union's FP7 ERC Starting Grant ``The dynamics of black holes: testing
the limits of Einstein's theory'' grant agreement No.~DyBHo--256667. A.C.~acknowledges funding received from the European Union's Horizon 2020 research and innovation programme under the Marie Sk\l odowska-Curie grant agreement No 655524.
M.R.~acknowledges partial financial support from the S\~ao Paulo Research Foundation (FAPESP), Grants No.~2013/09357-9 and No.~2015/14077-0. M.R.~is also grateful to S.W.~and the University of Nottingham for hospitality while this work was being completed. S.W.~acknowledges financial support provided under the Royal Society University Research Fellow (UF120112), the Nottingham Advanced Research Fellow (A2RHS2) and the Royal Society Project (RG130377) grants.
This research was supported in part by the Perimeter Institute for
Theoretical Physics. Research at Perimeter Institute is supported by
the Government of Canada through Industry Canada and by the Province
of Ontario through the Ministry of Economic Development $\&$
Innovation.
This project has received funding from the European Union's Horizon 2020 research and innovation programme under the Marie Sklodowska-Curie grant agreement No 690904.
%

\newpage

\onecolumngrid  \vspace{1cm} 
\begin{center}  
{\Large\bf Supplemental Material} 
\end{center} 
\appendix 
\tableofcontents  

\setcounter{section}{19}
\setcounter{equation}{0}

\section{Perturbations in viscous fluids}
\label{app_A}

In the following we shall derive the equations of motion for sound and surface waves in irrotational fluid flows. In principle, we need to take into account dissipative effects due to viscosity if we want our models to be as realistic as possible. However, as we shall see, viscosity effects are negligible for the superradiant phenomena we study in this letter. 

The dynamics of the fluid is governed by the continuity equation
\be \label{eq:cont}
\p_t \rho +\nabla \cdot \left(\rho \mathbf{v} \right)=0, 
\ee 
together with the Navier-Stokes equation~\cite{Landau:book_fluids},
\be \label{Eq:Navier-Stokes}
\rho D_t \mathbf{v}  = - \nabla P + \mu \nabla^2 
\mathbf{v} + \left( \xi + \frac{\mu}{3} \right) \nabla \left(\nabla \cdot \mathbf{v} \right),
\ee
where the fluid following derivative is given by
$D_t := \left(\p_t + \mathbf{v} \cdot \nabla \right)$.
Here $\rho$ is the density of the fluid, $\mathbf{v}$ is the velocity field of the flow, and $P=P_\mathrm{d}+P_\mathrm{s}$ is the total pressure, consisting of $P_\mathrm{d}$ the dynamic and $P_\mathrm{s}=\rho g z$ the hydrostatic pressures. Dissipative effects are accounted for by $\mu$ the dynamic and $\xi$ the second (bulk) viscosities of the fluid.

The above formulation of the Navier-Stokes equation holds for compressible flows, the only assumption required is that $\mu$ and $\xi$ are constant throughout the flow. The vector Laplacian of the velocity, by definition, is given by 
\be
\nabla^2 \mathbf{v} = \nabla \left( \nabla \cdot \mathbf{v} \right) - \nabla \times \left( \nabla \times \mathbf{v}\right), 
\ee
while the convective acceleration is
\be
\mathbf{v} \cdot \nabla \mathbf{v} = \frac{1}{2} \nabla \mathbf {v} ^2  + \left( \nabla \times \mathbf{v}\right)\times \mathbf{v}.
\ee

We assume that the flow is irrotational $ \nabla \times \mathbf{v} = 0$, which allows us to write the velocity in terms of a velocity potential $\psi$, so that $\mathbf{v} = \nabla \psi$. We shall also assume a barotropic fluid ($P$ depends only on $\rho$). This allows us to define the specific enthalpy $H(P)=\int_0^P dP' / \rho$ so that $\nabla H = \nabla P / \rho$. Consequently, the Navier-Stokes equation reads
\be 
\label{eq:ns}
\p_t \mathbf{\nabla} \psi +\frac{1}{2} \nabla (\nabla \psi \cdot \nabla \Phi) = - \nabla H + \beta \nabla \left(\nabla^2 \psi \right),
\ee
where $\beta := \xi/\rho + 4\nu/3$ incorporates both viscosity coefficients.
\subsection{Sound waves in viscous fluids}
\label{SoundWaves_App}

In order to investigate the propagation of sound waves in such a system, we consider the linear perturbation of the background flow. 
We assume an incompressible background flow with zero velocity. 
We write the total density, pressure, enthalpy, velocity potential, and velocity field as the sum of the background part plus perturbations ($\varepsilon$ is assumed to be a small parameter):
\be
\rho_{\rm tot} =\rho_0 + \varepsilon \rho_1, \qquad
P_{\rm tot} = P_0 + \varepsilon P_1, \qquad H_{\rm tot} = H_0 +\varepsilon H_1, \qquad \psi_{\rm tot} = \psi_0 + \varepsilon \psi_1, \qquad \mathbf{v}_{\rm tot} = \varepsilon \mathbf{v}_1.
\ee
From the continuity and the Navier-Stokes equations, we obtain the following equations for the perturbed quantities (terms of order $\varepsilon ^2$ and higher are neglected):
\begin{align} \label{eq:cont_pert}
\p_t \rho_1 + \rho_0 \nabla^2 \psi_1 =0 , \\
\nabla \widetilde{\p_t} \psi_1 + \nabla H_1 =0  ,  \label{eq:ns_pert}
\end{align}
where $\widetilde{\p_t} :=  (\p_t -  \beta \nabla^2)$ is an effective time derivative for the viscous fluid.
Note that Eq.~\eqref{eq:ns_pert} above is a total derivative. After integration, it becomes
\be
\widetilde{\p_t} \psi_1 = -H_1 = - \frac{P_1}{\rho} = - c^2 \frac{\rho_1}{\rho_0} ,  \label{eq:ns_Bernoulli-like}
\ee
where the speed of sound $c$ is defined through $c^2 = \left(\p P/ \p \rho \right)_S$.
Because of its simple form, which gives $\rho_1$ in terms of $\psi_1$, Eq.~\eqref{eq:ns_Bernoulli-like} can be used to eliminate $\rho_1$ in Eq.~\eqref{eq:cont_pert}, yielding the propagation equation for sound waves:
\be
\left( \p_t \widetilde \p_t  
-c^2    \nabla^2 \right) \psi_1   = 0. \label{eq:master} 
\ee
If there is no explicit time dependence in the parameters, we have $\p_t \psi_1 = -i \om \psi _1$, and therefore
\be
\nabla^2 \psi_1 + \frac{\omega^2}{c^2-i \omega \beta} \psi_1 = 0. \label{eq:master_0} 
\ee
Note that viscosity manifests itself through an imaginary term in the wave equation. When solving Eq.~\eqref{eq:master_0}, it is convenient to define an effective complex speed of sound $c_{\rm eff}^2 = c^2 - i\beta \omega$, which includes the viscous term. 

Since both the wave equation and the boundary (i.e.~the cylinder) are invariant under cylindrical symmetry, this equation can be reduced to an ordinary wave equation. Indeed, with the \textit{ansatz} $\psi_{1}(r,\phi, z) = (\varphi(r)/\sqrt{r}) e^{i m \phi - i \om t },$
eq.~\eqref{eq:master_0} reduces to
\be 
\p_r^2 \varphi + \left[\frac{\omega^2}{c_{\rm eff}^2} - \frac{1}{r^2}\left(m^2 - \frac{1}{4}\right) \right] \varphi = 0.  \label{eq:radial} 
\ee 

For sound waves in water, the typical wave speed is $c \sim 1500\,{\rm m}\, {\rm s}^{-1}$, while $\beta \sim 10^{-6} \,{\rm m}^{2}\, {\rm s}^{-1}$. Hence, for the purposes of this letter, viscosity effects can be neglected since they would become important only at very high frequencies $\omega \sim 10^{12} \, {\rm s}^{-1}$.

\subsection{Surface waves in viscous fluids}

Besides sound waves, another interesting framework for studying superradiance consists in the propagation of waves at 
the free surface of a fluid. More specifically, we consider a 3-dimensional water tank with a flat bottom. It is convenient to separate all quantities into a tangential and a parallel part. 
We write $(x_\parallel,z)$ for the spatial coordinates, where the $z$-direction is parallel to the local gravitational field, and we  choose $z = 0$ as the location of the flat bottom. Our objective is to compute the free surface height $z=h(t,x_\parallel)$. For this we assume an irrotational ($\mathbf {v} =  \nabla \psi$) and incompressible ($\nabla \cdot \mathbf {v} = 0$) flow. 
Since we are not interested in density perturbations of the fluid, 
the perturbations are also incompressible, i.e.~$\nabla \cdot \mathbf{v}_1 = 0$.

Under these assumptions, the velocity potential obeys the Laplace equation 
\be \label{Laplace}
\nabla ^2 \psi = 0.
\ee
 Consequently, the viscous term in Eq.~\eqref{eq:ns} is identically zero. At the surface $z=h(t,x_\parallel)$, after integration, Eq.~\eqref{eq:ns} reads
\be \label{BCT1}
\p_t \psi + \frac12 (\vec \nabla_{\parallel} \psi)^2 + \frac12 ( \partial_{z} \psi)^2 + g [h(t,x_\parallel)-h_0] + S(t,x_\parallel) = 0,
\ee
where we have assumed that, at spatial infinity, $h(t,x_\parallel) \rightarrow h_0 = \rm{constant}$ and $\nabla \psi = \mathbf{v} \rightarrow  0$. (If the boundary conditions are imposed at a finite radius instead of infinity, Eq.~\eqref{BCT1} picks up new constant terms which are not relevant for later steps). The term $S(t,x_\parallel)$ accounts for the normal stress boundary condition at the surface [see Eq.~(B.4) of \cite{Johnson}]:
\be
S(t,x_\parallel) = 2 \mu  \partial_{z} ^2 \psi,
\ee
plus non-linear terms which vanish for the background flow. All the derivatives of the velocity field must be evaluated at $z=h(t,x_\parallel)$. 

  On the bottom, due to the no-slip boundary condition, the velocity must be zero. On the free surface, a point dragged by the flow must stay on the surface. This means
\be
\p_z \psi - \vec \nabla h \cdot \vec \nabla \psi - \p_t h = 0,\label{BCT2}
\ee
where the field derivatives are, again, evaluated at $z=h(t,x_\parallel)$.
To relate $\p_z \psi$ to the tangential derivatives, we must integrate the equation of motion inside the tank. We write Eq.~\eqref{Laplace} as $\p_z^2 \psi = - \nabla_{\parallel} \psi$ and integrate it from the bottom to the free surface: 
\be
(\p_z \psi)_{z = h(t,x_\parallel)} - (\p_z \psi)_{z = 0} = \int_0^{h(t,x_\parallel)} \nabla_{\parallel} \psi \; dz. 
\ee
Since the flow cannot penetrate the bottom, we have $(\p_z \psi)_{z = 0} = 0$. Moreover, in the shallow water regime, i.e. when $h$ is much smaller than the typical wavelength, the integral on the right hand side of the above equation further simplifies, and we obtain 
\be \label{eq:shallow}
(\p_z \phi)_{z = h(t,x_\parallel)} \simeq - h(t,x_\parallel) \nabla_{\parallel} \psi. 
\ee
 We now consider the propagation of linear fluctuations. We assume that there is no background flow, and that the unperturbed surface is flat. Let
\beq
\psi(t,x_\parallel) &=&\psi_0 + \varepsilon \psi_1(t,x_\parallel),\\
h(t,x_\parallel) &=& h_0 + \varepsilon h_1(t,x_\parallel),
\eeq
where $\psi_0$ and $h_0$ are constants, and $\varepsilon$ once more is a small parameter. At linear order in $\varepsilon$ we have, from \eqref{BCT1}, 
\be \label{h1eq}
\p_t \psi_1 + g h_1 - 2 \nu\nabla_{\parallel}^2 \psi_1=0,
\ee
and, from \eqref{BCT2} and \eqref{eq:shallow},  
\be
-h_0 \nabla_{\parallel}^2 \psi_1 = \p_t h_1.
\ee
By eliminating $h_1$ from the equations above, one finds
\be \label{gwaeq}
\p_t ^2 \psi_1 - gh_0 \nabla_{\parallel}^2\psi_1 - 2 \nu\nabla_{\parallel}^2 \p_t \psi_1 = 0. 
\ee
With the \textit{ansatz} $\psi_1(t,r,\phi, z) = (\varphi(r)/\sqrt{r})  e^{-i\omega t + i m \phi}$,
this equation reduces to \eqref{eq:radial}, with 
$
c_{\rm eff}^2= c^2 - i \om \beta$, $c^2= g h_0$ and $ \beta = 2 \nu$.
As in the case of sound waves, viscosity will be non-negligible only at very high frequencies. Assuming a typical surface wave speed of $c \sim 1\,{\rm m}\, {\rm s}^{-1}$, this will occur around $\om \sim 10^6 \,  {\rm s}^{-1}$.

\section{Boundary layers and boundary conditions}
\label{BC_App}

Due to viscosity, the background flow must satisfy a no-slip boundary condition at the surface of the cylinder. If the velocity profile $\vec v = v(t,r) \hat \phi$ of the fluid and its pressure $P=P(t,r)$ are rotationally symmetric at all times, the Navier-Stokes equation reduces to 
\be \label{heateq}
\p_t v = \nu \p_r\left(\frac{1}{r}\p_r(rv)\right) 
= \nu \left(\p_r^2 v + \frac{\p_r v}{r} - \frac{v}{r^2} \right),
\ee
where $\nu=\mu/\rho_0$ is the kinematic viscosity and the pressure is determined by the relation $\p_r P = \rho_0 v^2 / r$. This is a heat equation, and describes the diffusion of angular momentum from the surface of the cylinder to the fluid. 
At early times, the fluid is essentially at rest, corresponding to the initial condition $v(0,r)=0$. Due to the no-slip boundary conditions, we also have $v(t,R_0)=\Omega R_0$ and $v(t,R_1)=0$. 
After equilibrium is reached, the angular velocity $v_{eq}(r)$ satisfying the boundary conditions can be determined analytically to be
\be \label{equilibrium_flow}
v_{eq}(r) = -\frac{\Om R_0^2}{R_{1}^2-R_0^2} r + \frac 1r \frac{\Om R_{1}^2 R_0^2}{R_{1}^2-R_0^2}.
\ee   

Since the superradiant effects we investigate in this letter arise only if there is a significant difference between the velocities of the fluid and of the cylinder, the experiment we propose should be performed at early times, before the whole fluid is set in motion. To be more precise, when the inner cylinder starts rotating, a boundary layer forms close to its surface. As long as the boundary layer is sufficiently smaller than the wavelength of the waves, the waves do not perceive it and, for all practical purposes, the fluid can be considered to be at rest. The timescale associated with the diffusion of angular momentum according to the heat equation is given by $\tau \sim (R_1-R_0)^2/\nu$. On the other hand, the typical timescale associated with a scattering experiment can be estimated as the time it takes for a wave to propagate from the outer cylinder to the inner one: $\tau ' \sim (R_1-R_0)/c$, where $c$ is the wave speed. As long as $\tau  \gg \tau '$, i.e.~$R_1 - R_0 \gg \nu / c$, the assumption of a negligible flow velocity is justified for scattering experiments.

\begin{figure}[t]
\begin{center}
\begin{tabular}{c}
\includegraphics[width=7cm]{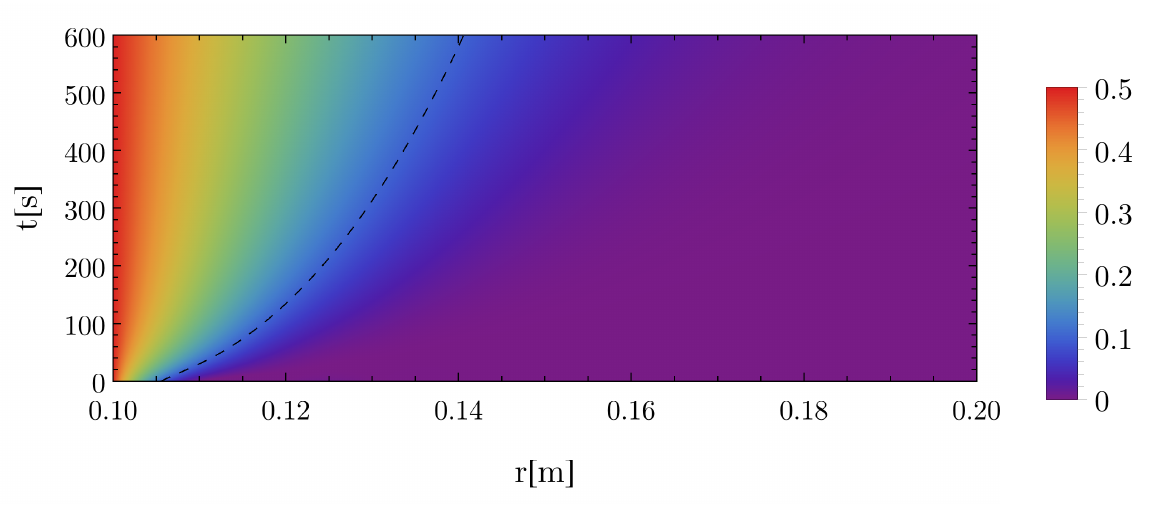}
\end{tabular}
\caption{Diffusion of angular momentum through water (at $20^{\circ}$\,C: $\nu \approx 10^{-6} \mathrm{m^2s^{-1}}$, $c \approx 1480 \mathrm{m s^{-1}}$) for $R_0=0.1\mathrm{m}$, $R_1=1\mathrm{m}$, $\Omega = 5 \mathrm{s^{-1}}$. The color axis represents the angular velocity $v$ as function of position and time. The dashed line corresponds to $v/(\Om R_0) = 0.2$.
\label{fig1}
}
\end{center}
\end{figure}

To estimate the diffusion of angular momentum, we consider, as an example, $R_0 \sim 0.1 \, \mathrm{m}$, $R_1 \sim 1 \, \mathrm{m}$ and $\Om \sim 5 \, \mathrm{s^{-1}}$. The corresponding numerical solution of the heat equation is plotted in Fig.~\ref{fig1}. After 600s, the fluid rotates with at least 20\% of the cylinder's velocity only inside a layer of $\sim 4 \, \mathrm{cm}$. This leaves enough time to perform the experiments here proposed. 

Another important detail regarding the boundary layer is its effect on the boundary conditions at the surface of the cylinders.
These boundary conditions are determined by the impedance of the cylinder, which is defined as the relation between the pressure change $P_1$ and the change of radial velocity $ \mathbf {v_1}$ at the surface of the cylinder (in the rest frame of the cylinder) through~\cite{Rienstra}
\be
Z_\om  =\mp \frac{P_1}{ \hat{r} \cdot \mathbf {v_1}},\label{impedance_BC}
\ee
where $\hat{r}$ is the unit vector in the radial direction. The upper signal is used for the inner cylinder and the lower one for the outer cylinder. To determine the correct boundary condition satisfied by the field, we must find the dynamical relation between the pressure change and the field. 

At early times, most of the fluid is at rest, and therefore, the velocity profile quickly drops from $\Om R_0$, the speed of the cylinder's surface, to zero over a short distance $\delta$. Within this boundary layer, the flow is nonzero and rotational, and relation \eqref{eq:ns_Bernoulli-like} cannot be used. However, if the flow in the boundary layer stays laminar, the pressure varies very few in the transverse direction, i.e.~$\p_r P_1 \approx 0$ (see e.g.~Ref~\cite{Landau:book_fluids}, Ch.~IV). This means that we can impose the boundary condition of the cylinder \eqref{impedance_BC} at $r = R_0 + \delta$, that is, at the edge of the boundary layer, where the flow becomes negligible ($\mathbf{v}_0 \approx 0$) and Eq.~\ref{eq:ns_Bernoulli-like} is valid again. From Eq.~\ref{eq:ns_Bernoulli-like}, we deduce that $P_1 = - \rho_0 \widetilde \p_t \psi_1$. Moreover, as we mentioned previously, unless one works at very high frequencies, viscosity can be neglected inside the fluid (but not inside the boundary layer), and therefore also around $r=R_0+\delta$. Hence $\widetilde \p_t$ can be safely replaced by $\p_t$ and we have $P_1 = - \rho_0 \p_t \psi_1$. 

Since the cylinder rotates uniformly with angular velocity $\Omega$, in order to use \eqref{impedance_BC} it is necessary to transform to a new angular coordinate, $\phi \rightarrow \phi+\Omega t$, which effectively amounts to the replacement $\widetilde \p_t \rightarrow \widetilde \p_t + \Om \partial_\phi$. Consequently, when cylindrical symmetry is taken into account, i.e. $\psi_1(t,r,\phi,z) = \psi_1(r) e^{-i \omega t + im\phi}$, this corresponds to the replacement $\om \rightarrow \tom = \om - m \Om$. 
Moreover, if the size of the boundary layer $\delta$ is much shorter than the typical wavelength $\lambda$, the value of $ \psi_1$ at $r=R_0+\delta$ is essentially that at $r=R_0$. This shows that the effect of the boundary layer can be neglected as long as $\delta \ll \lambda$ (this will ultimately break down due to diffusion of angular momentum in the fluid). 
In view of the above,
the boundary condition~\eqref{impedance_BC} at the surface of a rotating cylinder, for a wave of frequency $\om$, yields 
\begin{equation}
\left(\frac{\p_r \psi_1}{\psi_1}\right)_{r=R_0}=\mp\frac{i\rho_0 \tom}{Z_\omega} . \label{BCsound2}
\end{equation}

For surface waves, the discussion of the effect of the boundary layer is similar. What differs is the way to relate pressure change to the wave itself. For this, we use the fact that pressure is constant at the surface ($P_{\rm atm}$ denotes the atmospheric pressure): 
\beq
P(h_0 + \varepsilon h_1) = P(h_0) + \varepsilon h_1 \p_z  P(h_0) = P_{\rm atm}.
\eeq
Hence, at order $O(\varepsilon)$,
\be \label{p1}
P_1 = P(h_0) - P_{\rm atm} = - h_1 \p_z P(h_0) .
\ee
By deriving the hydrostatic pressure with respect to $z$, we see that $\p_z P = - \rho g$. This is the hydrostatic equilibrium. Together with Eqs.~\eqref{h1eq} and \eqref{p1}, we obtain 
\be
P_1 = - \rho_0 \p_t \psi_1 ,
\ee
which, combined with Eqs.~\eqref{gwaeq} and \eqref{impedance_BC}, produces the same boundary condition \eqref{BCsound2}.

\bibliographystyle{h-physrev4}
\bibliography{superradiance_ref}

\end{document}